\documentclass[journal]{IEEEtran}
\usepackage[pdftex]{graphicx}
\usepackage{amssymb}
\usepackage{epstopdf}

\usepackage{xcolor}
\usepackage{algorithm,algorithmic}

\usepackage[cmex10]{amsmath}

\ifCLASSINFOpdf
\else
\fi
\usepackage[cmex10]{amsmath}

\DeclareMathOperator\erfi{erfi}

\begin{document}
%
\title{Signal Space Cognitive Cooperation }
%
%
\author{Eylem Erdogan, Ali Afana, Hamza Umit Sokun, Salama Ikki, Lutfiye Durak-Ata and Halim Yanikomeroglu}
\author{Eylem~Erdogan,~\IEEEmembership{Member,~IEEE},~Ali~Afana,~\IEEEmembership{Member,~IEEE},~Hamza~Umit~Sokun,~\IEEEmembership{Member,~IEEE}, ~Salama~Ikki,~\IEEEmembership{Member,~IEEE},~Lutfiye~Durak-Ata,~\IEEEmembership{Senior~Member,~IEEE},~and~Halim~Yanikomeroglu,~\IEEEmembership{Fellow,~IEEE}.
\thanks{E. Erdogan is with the Department of Electrical and Electronics Engineering, Istanbul Medeniyet University, Uskudar, Istanbul, Turkey (e-mail: eylem.erdogan@medeniyet.edu.tr)} 
\thanks{A. Afana and S. S. Ikki are with the Department of Electrical Engineering, Lakehead University, Thunder Bay, ON, Canada (e-mail: \{aafana, sikki\}@lakeheadu.ca)}
\thanks{L. Durak-Ata is with the Informatics Institute, Istanbul Technical University, Maslak, Istanbul, Turkey. (e-mail: durakata@itu.edu.tr)}
\thanks{H. U. Sokun and H. Yanikomeroglu are with the Systems and Computer Engineering, Carleton University, Ottawa, ON, Canada (e-mail: \{husokun, halim\}@sce.carleton.ca)}

}

\maketitle


\begin{abstract}
In this work, a new transmission scheme, signal space cognitive cooperation, is introduced by applying the idea of signal space diversity in an underlay spectrum sharing decode-and-forward multi-relay cooperative network. In the proposed structure, the secondary source signal is rotated by a certain angle and then the source and the secondary best relay transmit the in-phase and the quadrature components of the rotated signal. As a consequence, two source signals, rather than one, are transmitted in two-time slots which improves data rates considerably, compared to the conventional cognitive cooperative schemes. In this work, proactive relaying mode is used in which the best relay is selected based on the max-min selection criterion before executing the transmission. Considering both statistical and the instantaneous channel state information of the feedback channel between the primary receiver and the secondary network, two power allocation methods are adopted at the source. For both methods, closed-form expressions of error probability are derived. Moreover, asymptotic analysis is performed and diversity gain is obtained to provide further insights about the system performance. Finally, analytical expressions are verified by Monte-Carlo simulations.

\end{abstract}




%
\IEEEpeerreviewmaketitle

\section{Introduction}

Several promising technologies have been developed to enhance the overall performance of wireless systems. Among these technologies, cognitive radio (CR)  and cooperative communications have recently been considered as two of the most important contributions to the progress in wireless systems \cite{Xiang}. In CR networks,  spectrum-sharing paradigm is proposed to enhance the overall spectrum efficiency. In particular, in the common underlay model, the secondary system is allowed to share the spectrum with the primary system subject to certain interference constraints \cite{Goldsmith}. On the other hand, cooperative technology is introduced to improve the  reliability, network coverage, and spectral efficiency for the future wireless communication systems (see \cite{Tao} and references therein). Thereby, it has been adopted by several recent standards, such as the 4G Long-Term Evolution (LTE) \cite{Dahlman}, and it is expected to be among the key enabling technologies in the upcoming 5G standards
 
The major drawback of cooperative communication networks is the need for orthogonal time/frequency slots to transmit data in the broadcasting and relaying phases of the relay-aided systems due to half duplex transmission. This reduces the overall system spectral efficiency and limits the achievable throughput. Hence, the goal of this work is to enhance the overall spectral efficiency by adopting signal space diversity (SSD) in cooperative CR systems. In SSD, constellation signals are rotated by a certain phase before the transmission. This rotation maps the original symbols to a new rotated in-phase ($I$) and quadrature ($Q$) components of the constellation signal. Then, an interleaver is used to provide independent channel fading coefficients of the $I$ and $Q$ components. Hence, spectral efficiency can be enhanced without any extra complexity \cite{Boutros}.

In the literature, SSD is adopted in cooperative systems in \cite{Ahmadzadeh}-\cite{Hamza2} and the references therein. In \cite{Ahmadzadeh}, a single-relay cooperative system is considered, where the source and the relay cooperate to send the complete message information to the destination. The performance analysis of the SSD cooperative system shows that it outperforms other cooperative schemes, such as distributed turbo-coded cooperative schemes and transmodulated structures \cite{Xie}. In \cite{Osama}, SSD is employed on a multi-relay decode-and-forward (DF) relaying system where error and outage probabilities are obtained for reactive and proactive relaying modes. Moreover, in  \cite{Hamza2}, the idea of SSD is used in a two-way DF relaying structure, where error probability analysis is performed. All the aforementioned works considered SSD in traditional cooperative relaying schemes, and to the best of the authors' knowledge, there is no previous work about SSD in cooperative CR systems
\\{\textbf{Motivation and contributions:}}
To fill in the gap, in this paper, a novel transmission scheme, signal space cognitive cooperation (SSCC) is introduced. SSCC combines underlay CR, best-relay selection and SSD-based proactive DF relaying. In this set-up, two symbols, rather than one symbol, are transmitted in two-time slots to enhance the spectral efficiency in link level considerably. The contribution of this paper is fourfold: $1)$ As the CR environment makes the system model more challenging due to the interference links, a new upper bound for the probability density function (pdf) of the end-to-end (e2e) SNR assuming both interference and maximum transmit power constraints, is introduced to simplify the error probability analysis; $2)$ closed-form pdf expressions are derived assuming two power allocation methods between primary receiver and secondary network, i.e., perfect and limited channel state information (CSI); $3)$ closed-form error probability expressions are derived for both approaches; $4)$ asymptotic analysis for error probability is performed to depict the impact of different system parameters on the performance of the proposed system. 

%

\begin{figure}[t]
\centering
\includegraphics[width=3.2in,height=1.4in]{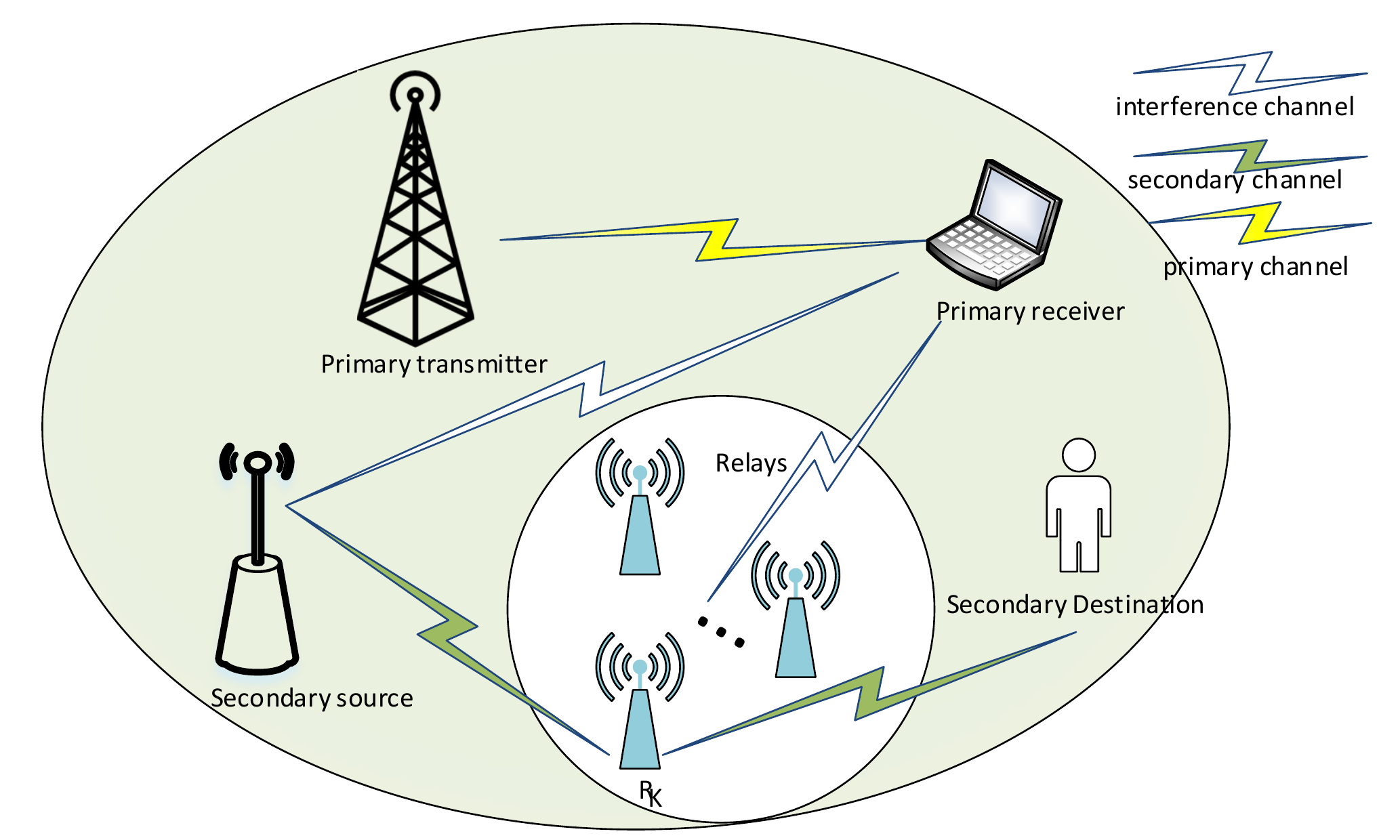}
\caption{System model of the proposed signal space cognitive cooperation.}
\label{fig_1}
\end{figure}

\section{Signal and System Model}

\subsection{System Model of the Secondary Network}

In this paper, a dual-hop underlay spectrum sharing DF cooperative network including of a primary receiver (PR) and secondary users{\footnote{{ It is important to note that the proposed scheme can be extended to various multi-user cognitive radio scenarios by using appropriate user scheduling approaches.}}}, is considered; see Fig. \ref{fig_1}. In the secondary network, source ($S$) intends to communicate with the destination ($D$) via the direct path and the $L$-th best ($R_L$) relay which is selected among $\mathcal{R}$ set of relays i.e., $\mathcal{R}=\{R_1,\cdots,R_L,\cdots,R_L\}$. To meet the interference power constraints of the PR, the transmit power at the secondary $S$ and $R_L$ is set to $P_{S}=\min\{\mathcal{Q}_P/|h_{S,P}|^2,P_{\max}\}$ and $P_{R}=\min\{\mathcal{Q}_P/|h_{R,P}|^2,P_{\max}\}$ \cite{Duong}, where $\mathcal{Q}_P$ is the maximum tolerable interference power at the PR and $P_{\max}$ is the total transmit power available in the network. As perfect CSI is assumed at $S \rightarrow P$ and $R_L \rightarrow P$, channel coefficients  $h_{S,P}$ and $h_{R_L,P}$ are modeled as complex Gaussian random variables with $(h_{S,P},h_{R_L,P})\sim\mathcal{C}\mathcal{N}(0,\sigma_{P}^2)$. Note that, primary transmitter is set to a far location from the secondary nodes and does not cause any interference on the secondary $R_L$ and $D$.

\subsection{Transmission with Signal Space Cognitive Cooperation}

In conventional cognitive cooperative systems, one source signal is transmitted in two-time slots due to half duplex relaying. However, by using the SSCC scheme, the number of signals that are transmitted in two-time slots can be doubled. This feature can be accomplished by rotating the original secondary source signal with a certain angle $\theta$ in which both the in-phase and the quadrature components of the rotated signal carry enough information to represent the original signal. To transmit the original signal components from $S$ to $D$, both the $S$ and the $R_L$ cooperate to transmit different copies of the rotated signal.

Let $X_S=\{X_1,X_2\}$ be a pair of the $S$ signals from the ordinary constellation $\mathcal{X}$ which are defined as $X_1=\Re\{X_1\}+j\Im\{X_1\}$ and $X_2=\Re\{X_2\}+j\Im\{X_2\}$, then rotated signals can be expressed as  $X_1^{\text{rot}}=\Re\{X_1^\text{rot}\}+j\Im\{X_1^\text{rot}\}$, $X_2^\text{rot}=\Re\{X_2^\text{rot}\}+j\Im\{X_2^\text{rot}\}$. Therefore, the signals that are transmitted from $S$ and $R_L$ can be formed by interleaving the in-phase and quadrature components of $X_1^\text{rot}$ and $X_2^\text{rot}$ as
\begin{align}
\lambda_s = \Re\{X_1^{\text{rot}}\}+j\Im\{X_2^{\text{rot}}\},\nonumber\\
\lambda_r = \Re\{X_2^{\text{rot}}\}+j\Im\{X_1^{\text{rot}}\},
\label{EQN:1}
\end{align}
where $\Re\{\cdot\}$ and $\Im\{\cdot\}$ denote the real and imaginary parts of the complex number and superscript $``\text{rot}"$ specifies the rotated signals. The transmission from $S$ to $D$ is completed in two-time slots. In the first time slot, $S$ transmits $\lambda_s$ to $R_L$ and $D$ with a signal power of $\min\{\mathcal{Q}_P/|h_{S,P}|^2,P_{\max}\}$. The received signal at the $R_L$ and $D$ can be written as 
\begin{align}
&y_{S,R_L} = \sqrt{\min\Big\{\frac{\mathcal{Q}_P}{|h_{S,P}|^2},P_{\max}\Big\}}h_{S,R_L}\lambda_s + n_R, \nonumber\\
&y_{S,D} = \sqrt{\min\Big\{\frac{\mathcal{Q}_P}{|h_{S,P}|^2},P_{\max}\Big\}}h_{S,D}\lambda_s + n_D.
\label{EQN:2}
\end{align}
In the second time slot, the $R_L$ detects the received signal and re-transmits it to the $D$. The received signal in the second phase can be written as 
\begin{align}
y_{R_L,D} = \sqrt{\min\Big\{\frac{\mathcal{Q}_P}{|h_{R_L,P}|^2},P_{\max}\Big\}}h_{R_L,D}\lambda_r + n_{R_L,D},
\label{EQN:3}
\end{align}
where $h_{S,R_L}$, $h_{S,D}$ and $h_{R_L,D}$ are modeled as $\mathcal{C}\mathcal{N}(0,\sigma_{S,R_L}^2)$, $\mathcal{C}\mathcal{N}(0,\sigma_{R_L,D}^2)$, and $\mathcal{C}\mathcal{N}(0,\sigma_{S,D}^2)$, respectively. Noise samples $n_R$, $n_D$ and $n_{R_L,D}$ are assumed to be complex additive white Gaussian noise (AWGN) samples which are modeled as zero-mean and unit-variance. To detect the original signals, $D$ has to reorder the received signals as
\begin{align}
  &\Delta_1=\Re\{h_{S,D}^*y_{S,D}\}, \nonumber\\
  &\Delta_2=\Im\{h_{S,D}^*y_{S,D}\}, \nonumber\\
  &\Delta_3=\Re\{h_{R_L,D}^*y_{R_L,D}\}, \nonumber\\
  &\Delta_4=\Im\{h_{R_L,D}^*y_{R_L,D}\},
  \label{EQN:4}
\end{align}
where $\{\Delta_1, \Delta_2, \Delta_3, \Delta_4\}$ are the reordered signals at the destination. Then, the maximum likelihood detection is applied to detect the source signals. The detailed transmission process of the proposed scheme proceeds in \textbf{Algorithm 1.} 

\begin{algorithm}[t]
\caption{Signal Space Cognitive Cooperation}
\begin{algorithmic}[1]
\algsetup{linenosize=\normalsize}
  \STATE Best relay $R_L$ is selected with the help of max-min selection criterion as given in (\ref{EQN:5}).
  \STATE   According to the modulation under consideration, the optimal rotation angle is obtained from Table I in \cite{Ahmadzadeh}. For instance, since we consider quadrature phase shift keying (QPSK) modulation in our work, the optimal value of $\theta$ is found as $26.6^\circ$ from Table I in \cite{Ahmadzadeh}. 
   \STATE By considering the interference channels between $S\rightarrow P$ and $R_L\rightarrow P$, the powers of $S$ and $R_L$ are set as $P_{S}=\min\{\mathcal{Q}_P/|h_{S,P}|^2,P_{\max}\}$ and $P_{R}=\min\{\mathcal{Q}_P/|h_{R,P}|^2,P_{\max}\}$.
  \STATE  In SSCC, the original source signals $X_1$ and $X_2$ are rotated by using optimum rotation angle $\theta$. By doing so, each point in the rotated constellation can be uniquely represented by real or imaginary parts. Thereby, in the first phase of the transmission, source signal can be formed by interleaving the in-phase and quadrature components of $X_1^\text{rot}$ and $X_2^\text{rot}$ as $\lambda_s = \Re\{X_1^{\text{rot}}\}+j\Im\{X_2^{\text{rot}}\}$. Then, $S$ transmits its information to $R_L$ and $D$.
  \STATE  In the second phase, the relay $R_L$ can easily decode $X_1^\text{rot}$ and $X_2^\text{rot}$ from $\Re\{X_1^{\text{rot}}\}$ and $\Im\{X_2^{\text{rot}}\}$ respectively. Then, $\lambda_r$ can be formed by interleaving the real and imaginary components of $X_1^{\text{rot}}$ and $X_2^{\text{rot}}$ as $\lambda_r = \Re\{X_2^{\text{rot}}\}+j\Im\{X_1^{\text{rot}}\}$ and then $R_L$ transmits $\lambda_r$ to the $D$.
  \STATE  $D$ combines the received signals from $S  \rightarrow D$ and $S \rightarrow R_L \rightarrow D$ by using maximum ratio combining. Then, the received signals are reordered as given in (4) and maximum likelihood detection is applied to detect the source signals.
  \STATE By combining spectrally efficient cognitive radio and relay aided transmission with signal space diversity, data rates can be doubled in the secondary networks without any additional complexity. Thereby, SSCC scheme can play a vital role in future wireless systems to enhance the overall spectral efficiency of the secondary network.
  \end{algorithmic}
\end{algorithm}

\subsection{End-to-End SNR Calculation}
In this paper, proactive relaying is used in which the best relay is selected based on the max-min selection criterion as
\begin{eqnarray}
R_L^* = \arg\max_{R_L\in\mathcal{R}}\min(\gamma_{S,R_L},\gamma_{R_L,D}), 
\label{EQN:5}
\end{eqnarray}
where  $\gamma_{S,R_L} = \min\Big\{\frac{\mathcal{Q}_P}{|h_{S,P}|^2},P_{\max}\Big\}|h_{S,R_L}|^2$ and $\gamma_{R_L,D} = \min\Big\{\frac{\mathcal{Q}_P}{|h_{R_L,P}|^2},P_{\max}\Big\}|h_{R_L,D}|^2$. The instantaneous SNRs between $S\rightarrow R_L\rightarrow D$ and $S\rightarrow D$ can be expressed as  
\begin{align}
&\gamma_{S,D}=\min\Big\{\frac{\mathcal{Q}_P}{|h_{S,P}|^2},P_{\max}\Big\}|h_{S,D}|^2,\nonumber\\ &\gamma_{S,R,D}=\smash{\displaystyle\max_{R_L\in\mathcal{R}}}\Big\{\min\Big(\gamma_{S,R_L},\gamma_{R_L,D}\Big)\Big\}. 
\label{EQN:6}
\end{align}
According to proactive relaying, end-to-end (e2e) SNR can be written as
\begin{align}
\gamma_{d} =
\gamma_{S,R,D}+\gamma_{S,D}.
\label{EQN:7}
\end{align}

\section{Performance Analysis}
In this section, the error probability performance of SSCC scheme is evaluated assuming perfect and limited feedback from the PR.

\subsection{SNR Statistics}
The cumulative distribution function (cdf) of $\gamma_{S,R,D}$ can be expressed as
\small
\begin{align}
F_{\gamma_{S,R,D}}&(\gamma)=\prod_{r=1}^{R_K}\Pr[\min(\gamma_{S,R_L},\gamma_{R_L,D})<\gamma] \nonumber\\ & =\prod_{r=1}^{R_K}\bigg\{1-(1-F_{\gamma_{S\rightarrow R_L}}(\gamma))(1-F_{\gamma_{R_L\rightarrow D}}(\gamma))\bigg\},
\label{EQN:8}
\end{align} 
\normalsize
whereas the cdf of $\gamma_{S,D}$  can be obtained  with the aid of \cite{Duong} as given in (\ref{EQN:9}) at the top of the next page.
\begin{figure*}[t]
\small
\begin{align}
&F_{\gamma_{S,D}}(\gamma)=F_{|{h_{S,P}}|^2}\bigg(\frac{{\mathcal{Q}}_p}{{P}_{\max}}\bigg)F_{|{h_{S,D}}|^2}\bigg(\frac{\gamma}{{P}_{\max}}\bigg)+\int_{\frac{{\mathcal{Q}}_p}{P_{\max}}}^{\infty}F_{|{h_{S,D}}|^2}\bigg(\frac{x\gamma}{\eta_{S,D}}\bigg)f_{|{h_{S,P}}|^2}(x)dx.
	\label{EQN:9}
\end{align}
\normalsize
\hrulefill
\end{figure*}
In (\ref{EQN:9}), $\eta_{S,D}=\frac{\sigma_{S,D}^2\mathcal{Q}_p}{\sigma_{P}^2}$. By inserting $F_{|{h_{S,D}}|^2}\bigg(\frac{x\gamma}{\eta_{S,D}}\bigg) = 1 - e^{\frac{x\gamma}{\eta_{S,D}}}$ and $f_{|{h_{S,P}}|^2}(x) = e^{-x}$ into (\ref{EQN:9}) and after few manipulations, $F_{\gamma_{S,D}}(\gamma)$ can be obtained as 
\begin{align}
F_{\gamma_{S,D}}(\gamma) = 1 + e^{-\frac{\gamma}{P_{\max}}}\Bigg(e^{-\frac{Q_p}{P_{\max}}}\Bigg(1-\frac{e^{-\frac{\gamma}{\eta_{S,D}}}}{1+\frac{\gamma}{\eta_{S,D}}}\Bigg)-1\Bigg).
	\label{EQN:10}
\end{align}
By substituting (\ref{EQN:10}) into (\ref{EQN:8}) and after replacing subscripts $\{S,D\}$ with $\{S,R_L\}$ and $\{R_L,D\}$, $F_{\gamma_{S,R,D}}(\gamma)$ can be  found. The pdf of ${\gamma_{S,R,D}}$ can be obtained by taking the derivative of $F_{\gamma_{S,R,D}}(\gamma)$ with respect to (w.r.t) $\gamma$ as seen in (\ref{EQN:11}) at the top of the next page. Note that, $\eta_{S,R_L}=\frac{\sigma_{S,R_L}^2\mathcal{Q}_p}{\sigma_{P}^2}$ and $\eta_{R_L,D}=\frac{\sigma_{R_L,D}^2\mathcal{Q}_p}{\sigma_{P}^2}$.
\begin{figure*}[t]
\small
\begin{align}
f_{\gamma_{S,R,D}}(\gamma)&=R_K\Bigg\{e^{-\frac{2\gamma}{P_{\max}}}\bigg(e^{-\frac{Q_p}{P_{\max}}}\bigg(1-\frac{e^{-\frac{\gamma}{\eta_{S,R_L}}}}{1+\frac{\gamma}{\eta_{S,R_L}}}\bigg)-1\bigg)\bigg(e^{-\frac{Q_p}{P_{\max}}}\bigg(1-\frac{e^{-\frac{\gamma}{\eta_{R_L,D}}}}{1+\frac{\gamma}{\eta_{R_L,D}}}\bigg)-1\bigg)\Bigg\}^{R_K-1}\nonumber\\&\times\Bigg[\frac{2e^{-\frac{2\gamma}{P_{\max}}}\bigg(e^{-\frac{Q_p}{P_{\max}}}\bigg(1-\frac{e^{-\frac{\gamma}{\eta_{S,R_L}}}}{1+\frac{\gamma}{\eta_{S,R_L}}}\bigg)-1\bigg)\bigg(e^{-\frac{Q_p}{P_{\max}}}\bigg(1-\frac{e^{-\frac{\gamma}{\eta_{R_L,D}}}}{1+\frac{\gamma}{\eta_{R_L,D}}}\bigg)-1\bigg)}{P_{\max}}\nonumber\\&-e^{-\frac{2\gamma+Q_p}{P_{\max}}}\bigg(\frac{e^{-\frac{\gamma}{\eta_{R_L,D}}}}{\big(1+\frac{\gamma}{\eta_{R_L,D}}\big)^2\eta_{R_L,D}}+\frac{e^{-\frac{\gamma}{\eta_{R_L,D}}}}{\big(1+\frac{\gamma}{\eta_{R_L,D}}\big)\eta_{R_L,D}}\bigg)\bigg(e^{-\frac{Q_p}{P_{\max}}}\bigg(1-\frac{e^{-\frac{\gamma}{\eta_{S,R_L}}}}{1+\frac{\gamma}{\eta_{S,R_L}}}\bigg)-1\bigg)\nonumber\\&-e^{-\frac{2\gamma+Q_p}{P_{\max}}}\bigg(\frac{e^{-\frac{\gamma}{\eta_{S,R_L}}}}{\big(1+\frac{\gamma}{\eta_{S,R_L}}\big)^2\eta_{S,R_L}}+\frac{e^{-\frac{\gamma}{\eta_{S,R_L}}}}{\big(1+\frac{\gamma}{\eta_{S,R_L}}\big)\eta_{S,R_L}}\bigg)\bigg(e^{-\frac{Q_p}{P_{\max}}}\bigg(1-\frac{e^{-\frac{\gamma}{\eta_{R_L,D}}}}{1+\frac{\gamma}{\eta_{R_L,D}}}\bigg)-1\bigg)\Bigg].
	\label{EQN:11}
\end{align} 
\hrulefill
\end{figure*}
\normalsize
Hence, by using (\ref{EQN:7}), the pdf of ${\gamma_{d}}$ can be obtained  as
\begin{eqnarray}
f_{\gamma_{d}}(\gamma) = f_{\gamma_{S,R_L,D}}(\gamma)\circledast f_{\gamma_{S,D}}(\gamma),
\label{EQN:12}	
\end{eqnarray}
where $\circledast$ shows the convolution operation. To the best of the authors' knowledge,  $f_{\gamma_{d}}(\gamma)$ cannot be obtained in closed-form. Nevertheless, after completing extensive simulation tests by using MATLAB program, a new upper bound on the pdf of $\gamma_{d}$ can be obtained as  
\begin{eqnarray}
f_{\gamma_{d}}(\gamma) \leq f_{\gamma_{up}}(\gamma)= f_{\gamma_{S,R_L,D}}(\gamma)f_{\gamma_{S,D}}(\gamma). 
\label{EQN:13}	
\end{eqnarray}
The upper bound on pdf proposed above eases the analysis and leads to a precise expression that enables us to evaluate the error performance without resorting to Monte Carlo simulations.  



\subsection{Average Error Probability}

Average error probability is an important performance indicator in wireless communications and it can be found as
\begin{align}
P_s(e)=\alpha\int_{0}^{\infty}Q\big(\sqrt{\beta\gamma}\big){f}_{\gamma_{up}}(\gamma)d\gamma,
\label{EQN:14}
\end{align}
where $\alpha$ and $\beta$ denote modulation coefficients \cite{Ahmadzadeh}. By substituting (\ref{EQN:13}) into (\ref{EQN:14}), average error probability can be derived. However, the result could not be obtained in closed form as the pdf of $\gamma_{up}$ is highly complicated. To solve this problem, we assume that all secondary relays are clustered together and experiencing the same scale fading \cite{Bodapati}, i.e., $\eta_{S,R_L}=\eta_{R_L,D} = \eta_R$. Moreover, to obtain a simple and tractable error probability expression, we assume that $S$ and $R_L$ are not power-limited terminals, i.e., $P_{\max}=\infty$ then, by applying Binomial approach and with the aid of well-known software programs like MATLAB or MATHEMATICA, $P_s(e)$ can be obtained as (\ref{EQN:15}) at the top of the next page. Note that, ${}_1 F_2(\cdot)$ denotes the generalized hypergeometric function \cite[eqn. 9.1]{Ryznik} and $\erfi(\cdot)$ stands for the imaginary error function.


\begin{figure*}[t!]
\small
\begin{align}
&P_s(e)= \alpha \sum_{r=0}^{R_K}\sum_{t=0}^{\infty}\binom{R_K}{r}\binom{2r+t+2}{t}(-1)^{n+t}\eta_{R}^{2r  + t + 2}\Bigg\{\frac{\eta_{S,D}^2\bigg(\Gamma\Big(2r-t+\frac{5}{2}\Big){}_p F_q\bigg(2,-2-2r+t;-2r+t-\frac{3}{2},t-2r-1;\frac{\beta\eta_{S,D}}{2}\bigg)}{{\sqrt{\pi}(2r-t+2)}\big(\frac{\beta}{2}\big)^{2r - t + 2}} \nonumber\\ &+ e^{\beta\eta_{S,D}/2}\sqrt{2\pi\beta}\eta_{S,D}^{2r-t+9/2}\sec(2\pi r - \pi t)- 2\pi\eta_{S,D}^{2r-t+4} (3+2r-t)\Bigg(\csc(2\pi r -  t) - \erfi\bigg(\sqrt{\beta\eta_{S,D}/2}\bigg)\sec(2\pi r -\pi t) \Bigg) \Bigg\},
	\label{EQN:15}
\end{align}
\hrulefill
\end{figure*}
\normalsize

\subsection{Impact of Limited Feedback}
In underlay CR networks, it is generally assumed that the instantaneous CSI of the interference channel or its gain, is known at the PR. Therefore, it can compute the mean-value (MV) of this gain and returns it back to the secondary source. Consequently, the adoption of the MV-power allocation can greatly minimize the feed-back burden in underlay CR networks \cite{Ali}. In the presence of MV power allocation, the transmit powers at the secondary $S$ and $R_L$ relay are set to $\hat{P}_S=\min\Big\{\mathcal{Q}_P/\mathbb{E}[|h_{S,P}|^2],P_{\max}\Big\}$ and  $\hat{P}_R=\min\Big\{\mathcal{Q}_P/\mathbb{E}[|h_{R_L,P}|^2],P_{\max}\Big\}$, respectively, where $\mathbb{E}[\cdot]$ is the expectation operator.



 By substituting $\hat{P}_S$ and $\hat{P}_R$ in (\ref{EQN:8}) and then by using  (\ref{EQN:13}), $\hat{f}_{\gamma_{up}}(\gamma)$ can be found as  
\small
\begin{align}
&\hat{f}_{\gamma_{up}}(\gamma) = \frac{R_K\mathcal{Z}}{\min\{\mathcal{Q}_P/\sigma_{P}^2,P_{\max}\}\sigma_{S,D}^2}\nonumber\\&\times\sum_{r=0}^{R_K-1}\binom{R_K-1}{r}(-1)^re^{-\gamma\bigg(\frac{(r+1)\mathcal{Z}\min\{\mathcal{Q}_P/\sigma_{P}^2,P_{\max}\}\sigma_{S,D}^2+1}{\min\{\mathcal{Q}_P/\sigma_{P}^2,P_{\max}\}\sigma_{S,D}^2}\bigg)},
	\label{EQN:16}
\end{align}
\normalsize
where $\mathcal{Z}=\frac{\min\big\{\mathcal{Q}_P/\sigma_P^2,P_{\max}\big\}\sigma_{S,R_L}^2+\min\big\{\mathcal{Q}_P/\sigma_P^2,P_{\max}\big\}\sigma_{R_L,D}^2}{\min\big\{\mathcal{Q}_P/\sigma_P^2,P_{\max}\big\}\sigma_{S,R_L}^2\min\big\{\mathcal{Q}_P/\sigma_P^2,P_{\max}\big\}\sigma_{R_L,D}^2}$. By substituting (\ref{EQN:16}) into (\ref{EQN:14}), the average error probability in the presence of limited feedback can be obtained as given in (\ref{EQN:17}) at the top of the next page.
\begin{figure*}
\small
\begin{align}
\hat{P}_s(e) = \frac{R_K\mathcal{Z}}{\min\{\mathcal{Q}_P/\sigma_{P}^2,P_{\max}\}\sigma_{S,D}^2}\sum_{r=0}^{R_K-1}\binom{R_K-1}{r}&(-1)^r\bigg\{2\bigg(\frac{(r+1)\mathcal{Z}\min\{\mathcal{Q}_P/\sigma_{P}^2,P_{\max}\}\sigma_{S,D}^2+1}{\min\{\mathcal{Q}_P/\sigma_{P}^2,P_{\max}\}\sigma_{S,D}^2}\bigg)\nonumber\\&+\beta\bigg[1+\sqrt{\bigg(\beta + 2\bigg(\frac{(r+1)\mathcal{Z}\min\{\mathcal{Q}_P/\sigma_{P}^2,P_{\max}\}\sigma_{S,D}^2+1}{\min\{\mathcal{Q}_P/\sigma_{P}^2,P_{\max}\}\sigma_{S,D}^2}\bigg)\bigg/\beta}\bigg]\bigg\}^{-1}.
	\label{EQN:17}
\end{align}
\hrulefill
\end{figure*}
\normalsize

\subsection{Analysis in the High SNR Regime}

At high SNR, $f_{\gamma_{up}}(\gamma)$ can be expressed as $f_{\gamma_{up}}^{\infty}(\gamma)\approx \{f_{\gamma_{S,R_L}}(\gamma)+f_{\gamma_{R_L,D}}(\gamma)\}^{R_K}f_{\gamma_{S,D}}(\gamma)$. By assuming $P_{\max}=\infty$, with the aid of Taylor expansion $(1+t)^{-a}=1-at+\ldots + \frac{a(a+1)\ldots(a+n-1)}{n!}t^n$, and after omitting the small-valued terms, $f_{\gamma_{up}}^{\infty(\gamma)}$ can be obtained as 
\begin{align}
f_{\gamma_{up}}^{\infty}(\gamma) \approx R_K\bigg(\frac{\kappa_1 + \kappa_2}{\kappa_1\kappa_2}\bigg)^{R_K}\frac{1}{\kappa_3}\bigg(\frac{1}{\bar{\gamma}}\bigg)^{R_K+1}\gamma^{R_K-1},
	\label{EQN:18}
\end{align}
where $\bar{\gamma}=\frac{\eta_{S,R_L}}{\kappa_1}=\frac{\eta_{R_L,D}}{\kappa_2}=\frac{\eta_{S,D}}{\kappa_3}$. By substituting (\ref{EQN:18}) into  (\ref{EQN:14}), asymptotic error probability can be obtained as
\begin{eqnarray}
P_s^{\infty}(e)=\frac{\alpha2^{R_K-1}\Gamma(R_K+1/2)}{\sqrt{\pi}\kappa_3}\bigg(\frac{\kappa_1 + \kappa_2}{\beta\kappa_1\kappa_2}\bigg)^{R_K}\bigg(\frac{1}{\bar{\gamma}}\bigg)^{\mathcal{G}_d}.
	\label{EQN:19}
\end{eqnarray}
The diversity order of the SSCC system can be found as  $\mathcal{G}_d = R_K+1$. 


\begin{figure}[t]
\centering
\includegraphics[width=3.2in]{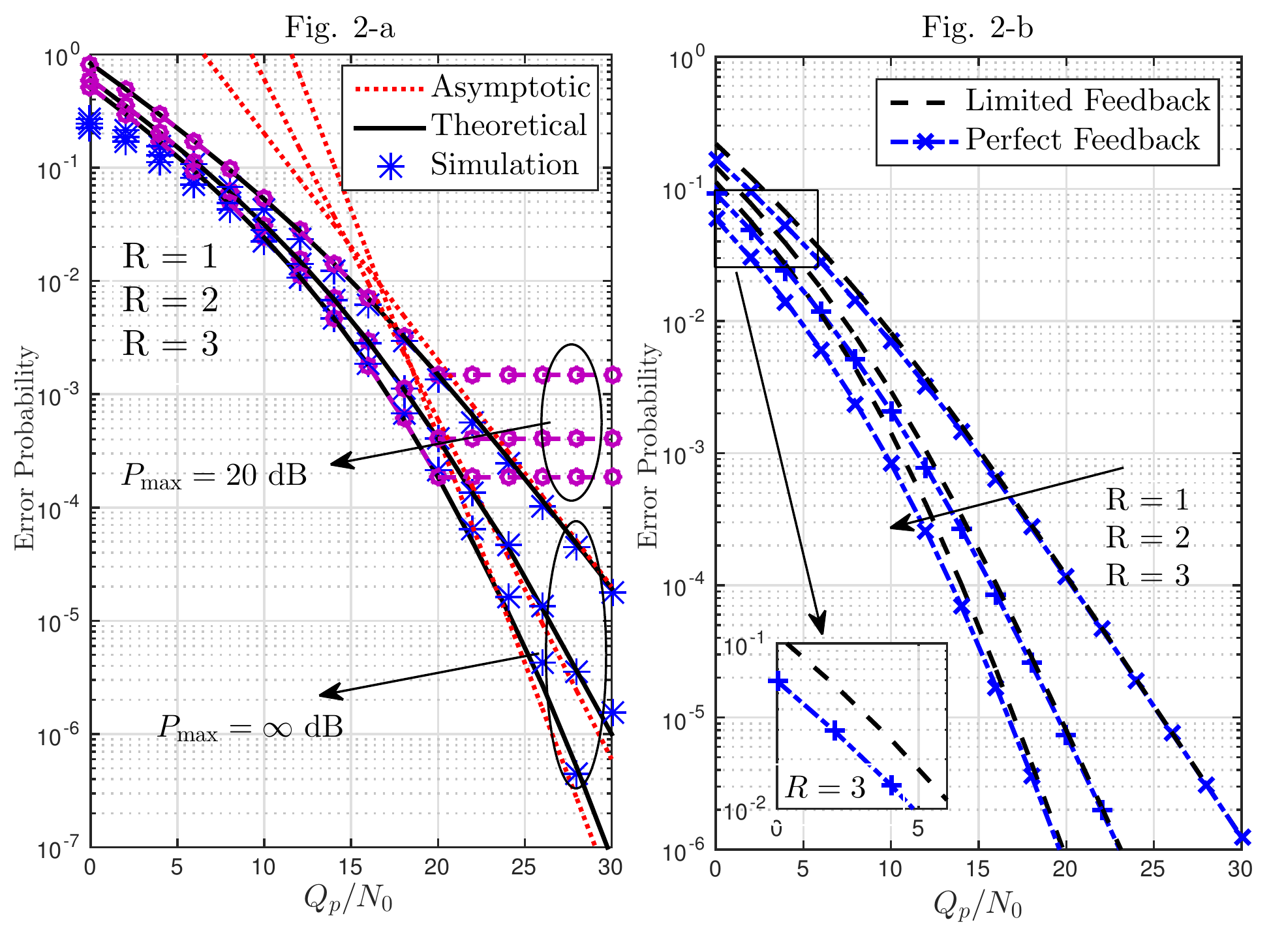}
\vspace{-0.4cm}
\caption{Error probability of the SSCC scheme for different number of relays.}
\label{fig_2}
\end{figure}
\begin{figure}[t]
\centering
\includegraphics[width=3.1in]{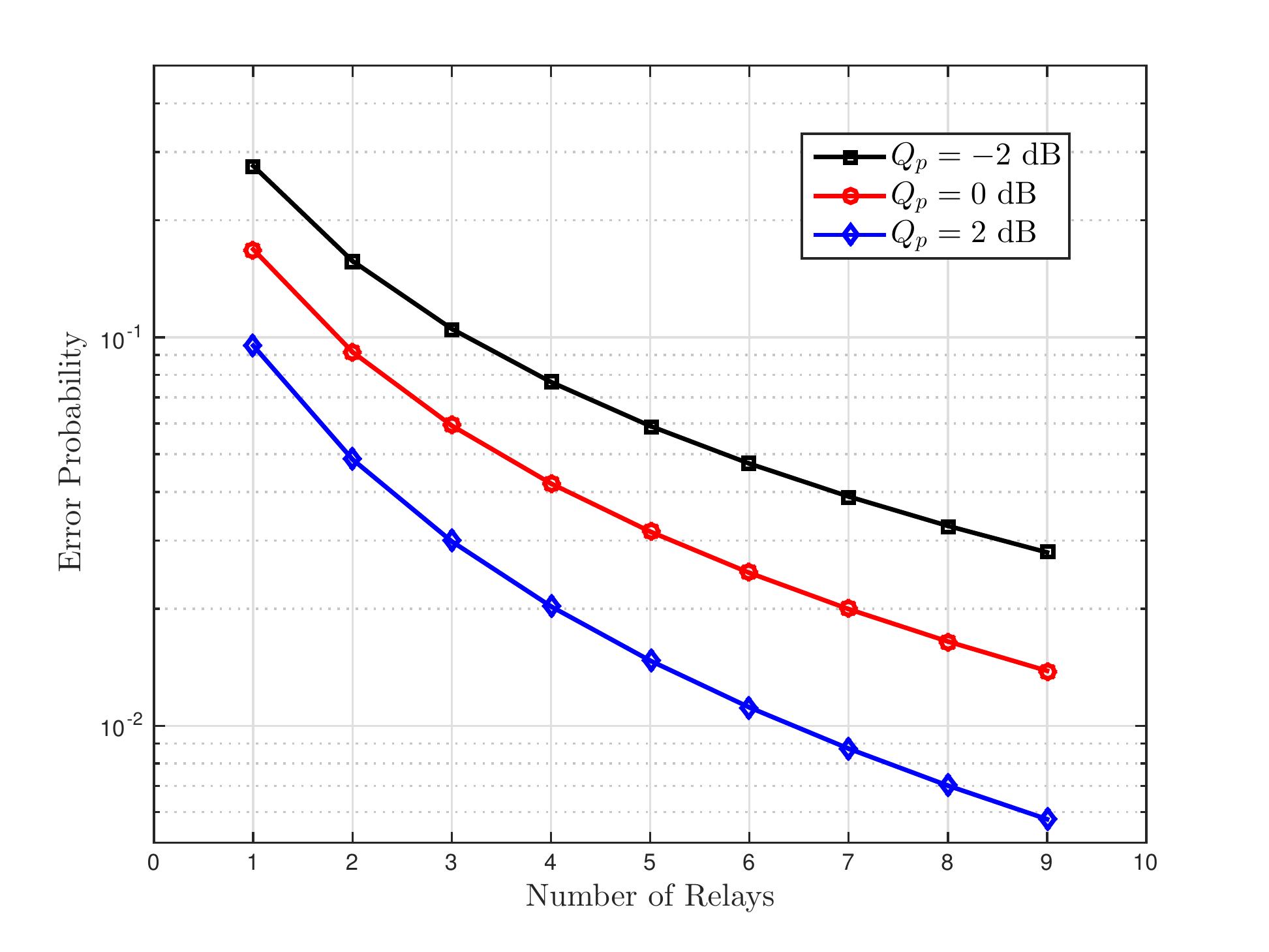}
\vspace{-0.4cm}
\caption{Error probability of the SSCC scheme for various power levels.}
\label{fig_3}
\end{figure}

\emph{\textbf{Remark}:} As seen from the analysis, SSCC scheme achieves full diversity available in the secondary network. Moreover, data rates can be doubled as the transmission between $S$ to $D$ is completed in one time slot. Note that, like SSCC, full-duplex relaying can achieve same data rates with a loss in the overall performance due to residual self-interference \cite{Challanges1}, \cite{Challanges2}. Thereby, SSCC can be a preferable scheme for next generation wireless systems.

\section{Numerical Results}

In this section, various numerical examples are provided to demonstrate the performance of SSCC. In the simulations, QPSK is considered with optimum rotation angle equal to $\theta = 26.6^\circ$ \cite{Ahmadzadeh}. 

{Fig. \ref{fig_2}}-a depicts the error probability performance of the SSCC scheme for different number of relays when $\sigma_{R}^2=\sigma_{P}^2=1$. As can be seen, the theoretical curves match with the simulations at especially medium and high SNRs, and as the total number of $R_K$ increases, the error performance improves substantially because of increasing number of available paths in the transmission. Moreover, purple colored dotted lines initially saturates due to the peak interference constraint. 

Fig. \ref{fig_2}-b compares perfect feedback power allocation with the limited feedback case for $\sigma_{R}^2= 4, \sigma_{P}^2=1$. The figure shows that limited feedback decreases the overall error probability performance at low and medium SNRs as only the mean-value of the feedback channel is available at the $S$. However, there is no performance loss at high SNR regime.

In Fig. \ref{fig_3}, maximum tolerable interference power of the primary system of the SSCC scheme is set to $Q_p = -2$ dB, $Q_p = 0$ dB, and $Q_p = 2$ dB. It can be observed from the figure that, increasing the number of relays do not monotonically improve the system performance. On the contrary, after a few number of relays, the error performance slowly saturates as the system performance reaches to its peak value. Thereby, selecting one of a few number of relays will lead to optimum error performance for the SSCC scheme.

\section{Conclusions}

In this work, a spectrally efficient transmission scheme called SSCC was introduced by combining underlay CR, best-relay selection, and SSD-based proactive DF relaying. In the analysis of the scheme, two power allocation approaches are considered at the secondary system, assuming instantaneous (perfect) and limited feedback from the primary receiver. Error probability and error asymptotic probability expressions are derived for both approaches using the proposed pdf upper bound. Results show that the SSCC scheme proposed herein can be a promising solution to improve spectral efficiency in future wireless systems.


\begin{thebibliography}{1}


\bibitem{Xiang}
W. Cheng-Xiang \textit{et al.}, ``Cellular architecture and key technologies for 5G wireless communication networks,''  \emph{ IEEE Commun. Mag.}, vol. 52, no. 2, pp. 122-130, Feb. 2014.

\bibitem{Goldsmith}
A. Goldsmith, S. A. Jafar, I. Maric, and S. Srinivasa, ``Breaking spectrum gridlock with cognitive radios: An information theoretic perspective,"  \emph{Proc. of the IEEE,} vol. 97, no. 5, pp. 894-914, May 2009.

\bibitem{Tao}
X. Tao, X. Xu, and Q. Cui, ``An overview of cooperative communications,''
\emph{IEEE Commun. Mag.}, vol. 50, no. 6, pp. 65-71, Jun. 2012.

\bibitem{Dahlman}
E. Dahlman, S. Parkvall, J. Skold, and P. Beming, ``3G evolution: HSPA
and LTE for mobile broadband,'' New York, Academic Press, 2010.

\bibitem{Boutros}
J. Boutros and E. Viterbo, ``Signal space diversity: A power-and
bandwidth-efficient diversity technique for the Rayleigh fading channel,''
\emph{IEEE Trans. Inf. Theory}, vol. 44, no. 4, pp. 1453-1467, Jul. 1998.

\bibitem{Ahmadzadeh}
 S. A. Ahmadzadeh, S. A. Motahari, and A. K. Khandani, ``Signal space
cooperative communication,'' \emph{IEEE Trans. Wireless Commun.}, vol. 9,
no. 4, pp. 1266-1271, Apr. 2010.


\bibitem{Xie}
Q. Xie, J. Song, K. Peng, F. Yang, and Z. Wang, ``Coded modulation with signal space diversity,'' \emph{IEEE Trans. Wireless Commun.}, vol. 10, no. 2,
pp. 660-669, Feb. 2011.


\bibitem{Osama}
O. Amin, R. Mesleh, S. Ikki, M. H. Ahmed, and O. A. Dobre, ``Performance analysis of multiple-relay cooperative systems with signal space diversity,'' \emph{IEEE Trans. on Vehic. Tech.}, vol. 64, no. 8, pp. 3414-3425, Aug. 2015.


\bibitem{Hamza2}
H. U. Sokun, M. C. Ilter, S. Ikki, and H. Yanikomeroglu, ``A spectrally efficient signal space diversity-based two-way relaying system,''  \emph{IEEE Trans. on Vehic. Tech.}, vol. 66, no.7, pp. 6215-6230, July 2017.  


\bibitem{Duong}
T. Q. Duong, D. B. da Costa, M. Elkashlan, and V. N. Q. Bao, ``Cognitive amplify-and-forward relay networks over Nakagami-m fading'', \emph{IEEE Trans. on Vehic. Tech.}, vol. 61, no. 5, pp. 2368-2374, June 2012.



\bibitem{Bodapati}
H. K. Boddapati, M. R. Bhatnaga and S. Prakriya, ``Performance analysis of cluster-based multi-hop underlay CRNs using max-link-selection protocol'', \emph{IEEE Trans. on Cogn. Commun. and Networks}, vol. 4, no. 1, pp. 15-29, Mar. 2018.

\bibitem{Ryznik}
I.~S.~Gradshteyn and I.~M.~Ryzhik, ``Table of integrals, series and products,'' 7th ed. New York: Academic Press, 2007.


\bibitem{Ali}
A.~Afana, T.~M.~N.~Ngatched, and O.~A.~Dobre, ``Spatial modulation in MIMO limited-feedback spectrum-sharing systems with mutual interference and channel estimation errors,'' \emph{IEEE Commun. Lett.}, vol. 19, no. 10, pp. 1754-1757, Aug. 2015.

\bibitem{Challanges1} 
T. Riihonen, S. Werner, and R. Wichman, ``Mitigation of loopback self-interference in full-duplex MIMO relays,'' \emph{IEEE Trans. on Signal Process.}, vol. 59, pp. 5983-5993, Dec. 2011. 

\bibitem{Challanges2} 
A. Sabharwal, \emph{et . al.,} ``In-band full-duplex wireless: Challenges and opportunities,'' \emph{IEEE J. Sel. Areas Commun.}, vol. 32, no. 9, pp. 1637-1652, Jun. 2014.

\end{thebibliography}
\end{document}